\documentclass[final,5p,times,twocolumn]{elsarticle}

\usepackage{ifthen} 
\newboolean{pdflatex}
\setboolean{pdflatex}{true} 

\newboolean{articletitles}
\setboolean{articletitles}{true} 

\newboolean{uprightparticles}
\setboolean{uprightparticles}{false} 

\newboolean{inbibliography}
\setboolean{inbibliography}{false} 

\def\lhcb {\mbox{LHCb}\xspace}
\ifthenelse{\boolean{uprightparticles}}%
{

 \def\Ppsi        {\ensuremath{\uppsi}\xspace}

 \def\PDelta      {\ensuremath{\Delta}\xspace}
 \def\PXi      {\ensuremath{\Xi}\xspace}
 \def\PLambda      {\ensuremath{\Lambda}\xspace}
 \def\PSigma      {\ensuremath{\Sigma}\xspace}
 \def\POmega      {\ensuremath{\Omega}\xspace}
 \def\PUpsilon      {\ensuremath{\Upsilon}\xspace}

 \def\PB      {\ensuremath{\mathrm{B}}\xspace}
 
 \def\PD      {\ensuremath{\mathrm{D}}\xspace}

 \def\PJ      {\ensuremath{\mathrm{J}}\xspace}
 \def\PK      {\ensuremath{\mathrm{K}}\xspace}

 \def\Pi      {\ensuremath{\mathrm{i}}\xspace}

 \def\Ps      {\ensuremath{\mathrm{s}}\xspace}

}
{

 \def\Ppsi        {\ensuremath{\psi}\xspace}
 
 \mathchardef\PDelta="7101
 \mathchardef\PXi="7104
 \mathchardef\PLambda="7103
 \mathchardef\PSigma="7106
 \mathchardef\POmega="710A
 \mathchardef\PUpsilon="7107
 
 \def\PB      {\ensuremath{B}\xspace}
 
 \def\PD      {\ensuremath{D}\xspace}

 \def\PJ      {\ensuremath{J}\xspace}
 \def\PK      {\ensuremath{K}\xspace}

 \def\Pi      {\ensuremath{i}\xspace}

 \def\Ps      {\ensuremath{s}\xspace}

}

\makeatletter
\ifcase \@ptsize \relax
  \newcommand{\miniscule}{\@setfontsize\miniscule{4}{5}}
\or
  \newcommand{\miniscule}{\@setfontsize\miniscule{5}{6}}
\or
  \newcommand{\miniscule}{\@setfontsize\miniscule{5}{6}}
\fi
\makeatother

\DeclareRobustCommand{\optbar}[1]{\shortstack{{\miniscule (\rule[.5ex]{1.25em}{.18mm})}
  \\ [-.7ex] $#1$}}

\def\squark    {{\ensuremath{\Ps}}\xspace}

  \def\Kbar    {{\kern 0.2em\overline{\kern -0.2em \PK}{}}\xspace}

\def\KorKbar    {\kern 0.18em\optbar{\kern -0.18em K}{}\xspace}

  \def\Dbar    {{\kern 0.2em\overline{\kern -0.2em \PD}{}}\xspace}
\def\D       {{\ensuremath{\PD}}\xspace}

\def\DorDbar    {\kern 0.18em\optbar{\kern -0.18em D}{}\xspace}
\def\Dz      {{\ensuremath{\D^0}}\xspace}

\def\B       {{\ensuremath{\PB}}\xspace}
\def\Bbar    {{\ensuremath{\kern 0.18em\overline{\kern -0.18em \PB}{}}}\xspace}

\def\BorBbar    {\kern 0.18em\optbar{\kern -0.18em B}{}\xspace}

\def\Bs      {{\ensuremath{\B^0_\squark}}\xspace}

\def\jpsi     {{\ensuremath{{\PJ\mskip -3mu/\mskip -2mu\Ppsi\mskip 2mu}}}\xspace}

  \def\Y#1S{\ensuremath{\PUpsilon{(#1S)}}\xspace}

\def\Lbar        {{\ensuremath{\kern 0.1em\overline{\kern -0.1em\PLambda}}}\xspace}
\def\LorLbar    {\kern 0.18em\optbar{\kern -0.18em \PLambda}{}\xspace}

\def\to                 {\ensuremath{\rightarrow}\xspace}

\def\AT#1     {\ensuremath{A_{\mathrm{T}}^{#1}}\xspace}           

\def\C#1      {\ensuremath{\mathcal{C}_{#1}}\xspace}                       
\def\Cp#1     {\ensuremath{\mathcal{C}_{#1}^{'}}\xspace}                    
\def\Ceff#1   {\ensuremath{\mathcal{C}_{#1}^{\mathrm{(eff)}}}\xspace}        
\def\Cpeff#1  {\ensuremath{\mathcal{C}_{#1}^{'\mathrm{(eff)}}}\xspace}       
\def\Ope#1    {\ensuremath{\mathcal{O}_{#1}}\xspace}                       
\def\Opep#1   {\ensuremath{\mathcal{O}_{#1}^{'}}\xspace}                    

\newcommand{\tev}{\ifthenelse{\boolean{inbibliography}}{\ensuremath{~T\kern -0.05em eV}\xspace}{\ensuremath{\mathrm{\,Te\kern -0.1em V}}}\xspace}
\newcommand{\gev}{\ensuremath{\mathrm{\,Ge\kern -0.1em V}}\xspace}
\newcommand{\mev}{\ensuremath{\mathrm{\,Me\kern -0.1em V}}\xspace}
\newcommand{\kev}{\ensuremath{\mathrm{\,ke\kern -0.1em V}}\xspace}
\newcommand{\ev}{\ensuremath{\mathrm{\,e\kern -0.1em V}}\xspace}
\newcommand{\gevc}{\ensuremath{{\mathrm{\,Ge\kern -0.1em V\!/}c}}\xspace}
\newcommand{\mevc}{\ensuremath{{\mathrm{\,Me\kern -0.1em V\!/}c}}\xspace}
\newcommand{\gevcc}{\ensuremath{{\mathrm{\,Ge\kern -0.1em V\!/}c^2}}\xspace}
\newcommand{\gevgevcccc}{\ensuremath{{\mathrm{\,Ge\kern -0.1em V^2\!/}c^4}}\xspace}
\newcommand{\mevcc}{\ensuremath{{\mathrm{\,Me\kern -0.1em V\!/}c^2}}\xspace}

\def\mhz  {\ensuremath{{\rm \,MHz}}\xspace}

\def\gsim{{~\raise.15em\hbox{$>$}\kern-.85em
          \lower.35em\hbox{$\sim$}~}\xspace}
\def\lsim{{~\raise.15em\hbox{$<$}\kern-.85em
          \lower.35em\hbox{$\sim$}~}\xspace}
\def\mysim{\ensuremath\sim\kern-0.3em\xspace}

\def\pt         {\mbox{$p_{\rm T}$}\xspace}

\def\tell1  {TELL1\xspace}
\def\ukl1   {UKL1\xspace}

\newcommand{\ie}{\mbox{\itshape i.e.}\xspace}

\usepackage{microtype}
\usepackage{lineno}
\usepackage{xspace}
\usepackage{caption}
\usepackage{comment}

\usepackage[utf8]{inputenc}

\usepackage{graphicx}  
\usepackage{color}
\usepackage{colortbl}
\graphicspath{{./figs/}} 
\graphicspath{{./figs/gpd_inputs/}} 

\usepackage{amsmath}
\usepackage{amssymb}
\usepackage{amsfonts}
\usepackage{upgreek}
\usepackage{listings}
\definecolor{dkgreen}{rgb}{0,0.6,0}
\definecolor{gray}{rgb}{0.5,0.5,0.5}
\definecolor{mauve}{rgb}{0.58,0,0.82}

\newcommand*\patchAmsMathEnvironmentForLineno[1]{%
\expandafter\let\csname old#1\expandafter\endcsname\csname #1\endcsname
\expandafter\let\csname oldend#1\expandafter\endcsname\csname
end#1\endcsname
 \renewenvironment{#1}%
   {\linenomath\csname old#1\endcsname}%
   {\csname oldend#1\endcsname\endlinenomath}%
}
\newcommand*\patchBothAmsMathEnvironmentsForLineno[1]{%
  \patchAmsMathEnvironmentForLineno{#1}%
  \patchAmsMathEnvironmentForLineno{#1*}%
}
\AtBeginDocument{%
\patchBothAmsMathEnvironmentsForLineno{equation}%
\patchBothAmsMathEnvironmentsForLineno{align}%
\patchBothAmsMathEnvironmentsForLineno{flalign}%
\patchBothAmsMathEnvironmentsForLineno{alignat}%
\patchBothAmsMathEnvironmentsForLineno{gather}%
\patchBothAmsMathEnvironmentsForLineno{multline}%
\patchBothAmsMathEnvironmentsForLineno{eqnarray}%
}

\usepackage{hyperref}    
\usepackage[all]{hypcap} 

\hypersetup{
    unicode=false,          
    pdftoolbar=true,        
    pdfmenubar=true,        
    pdffitwindow=false,     
    pdfstartview={FitH},    
    pdfauthor={Sean Benson, Konstantin Gizdov},     
    pdfcreator={Sean Benson, Konstantin Gizdov},   
    pdfnewwindow=true,      
    colorlinks=true,       
    linkcolor=blue,          
    citecolor=green,        
    filecolor=magenta,      
    urlcolor=cyan           
}

\begin{document}

\begin{frontmatter}
\author[nl]{Sean Benson}
\ead{sean.benson@nikhef.nl}
\author[uoe]{Konstantin Gizdov}
\address[nl]{Nikhef National Institute for Subatomic Physics, Amsterdam, The Netherlands}
\address[uoe]{School of Physics and Astronomy, University of Edinburgh, Edinburgh, United Kingdom}
\title{NNDrone : a toolkit for the mass application of machine learning in High Energy Physics}
\begin{abstract}
  \noindent
  Machine learning has proven to be an indispensable tool in the selection of
  interesting events in high energy physics. Such technologies will become increasingly
  important as detector upgrades are introduced and data rates increase by orders of magnitude.
  We propose a toolkit to enable the creation of a drone classifier from any machine learning
  classifier, such that different classifiers may be standardised into a single form and executed
  in parallel. We demonstrate the capability of the drone neural network
  to learn the required properties of the input neural network without
  the use of any labels from the training data, only using appropriate questioning of the input neural network.
\end{abstract}
\end{frontmatter}

\section{Introduction}
\label{sec:intro}

Data-collection rates in high energy physics (HEP) experiments, particularly those at the Large Hadron Collider (LHC),
are a continuing challenge and resulting datasets require large amounts of computing power to process.
For example, the \lhcb experiment~\cite{Alves:2008zz} processes an event rate of 1\mhz in a software-based
trigger~\cite{LHCb-DP-2014-002}. The purpose of this trigger is to reduce the output
data rate to manageable levels, \ie to fit in the available storage resources offline.
This amounts to a reduction from 60\,GB per second to an output data rate of 0.6\,GB per second.
In order to accomplish such a remarkable real-time data reduction in the software based trigger,
novel ideas have been introduced, such as the real-time alignment and calibration of the detector~\cite{Xu:2016mik},
in addition to the concept of real-time analysis~\cite{Aaij:2016rxn}, whereby a subset of the particles from the proton collisions need only
be saved, and not the raw data from the sub-detectors.
The aforementioned data-reduction strategy is similar across all LHC experiments, where
software based selections are applied in low-latency environments.

Machine learning (ML) is becoming an increasingly important tool to filter datasets,
be it with the identification of interesting event topologies, or the distinction
between individual particle species. For the case of \lhcb data-taking, over 600
unique signatures are searched for in parallel in real time, each with its own set of requirements.
However only a handful at present make use of machine learning.

A large ecosystem is available for analysts to create machine learning classifiers;
the TMVA~\cite{Hocker:2007ht} and Neurobayes~\cite{Feindt:2006pm} tools being among the most widely used.
More recent examples gaining popularity include Scikit-Learn~\cite{Pedregosa:2012toh}
and Keras~\cite{keras}. It has been proven in many LHC analyses that
ML classifiers account for differences in the correlations of
training variables between signal and background events, therefore enabling more
powerful data reduction.
Despite this, the majority of searches for interesting signatures are performed
without the use of ML classifiers. Often the reason for this is the relative difficulty in
the implementation of a preferred ML classifier to the {\tt C++/Python} combination
of event selection frameworks~\cite{Barrand:2001ny}. Another
reason is the required algorithm speed. Methods such as Bonsai
Boosted Decision Trees (BBDTs)~\cite{Gligorov:2012qt} have been developed in order
to enable the quick evaluation of models. The BBDT approach relies on the
discretization of inputs such that all possible combinations along with
the associated classifier response is known before the model is evaluated.
One potential drawback of the BBDT approach is that the number of input variables is limited
in order to limit the number of possible combinations.

We present in this article a package that allows an analyst to
train a drone neural network that learns the important features of a
given ML learning classifier from any chosen package such as SciKit-Learn or Keras.
The resulting parameters are then fed into a {\tt C++} algorithm that
performs execution in HEP production environments. The details of the
drone training are provided in Sec.~\ref{sec:dlearn}. This is followed
by real examples using simulated data in Sec.~\ref{sec:hep}. The advantages
of the approach are discussed in Sec.~\ref{sec:storage} and a summary is
provided in Sec.~\ref{sec:summary}.

\section{Drone learning}
\label{sec:dlearn}

The training of the drone network requires that the original network is
extensively probed in the parameter space in which accuracy is desired.
The principle utilised in the training of the drone is that sufficient
approximation of the original network is achieved with sufficient expansion
of the hyperparameter space of the drone, and that the same global minimum
of the loss function can be found, as reported in Ref.~\cite{losssurfaces}.
The ability of a neural network with a continuous, bounded, non-constant activation
function to approximate functions to an arbitrary degree has been indeed known
since the early 1990s~\cite{HORNIK1991251}.

\subsection{Initial drone structure and corresponding training}

The drone chosen for use in this article is initialised as a
neural network with a single intermediate (hidden) layer of 5
nodes using a standard sigmoid activation function. The network
has the number of inputs determined from the number of desired
characteristics of the decay signature. A single output is taken
from the network and a linear model is used to relate layers.

The model is made to approximate the original classifier through
a supervised learning technique, though not in the traditional sense.
Instead of a label as {\tt signal} or {\tt background} taken from the training data, the
output of the original classifier is used as a label. This means that the
loss function is defined as
\begin{align}
\mathcal{L} = \sum_i \left( F(\vec{x}_i) - G_i(\vec{x}_i) \right)^2,
\end{align}
where $F(\vec{x}_i)$ and $G(\vec{x}_i)$ are the outputs
of the original and drone models on datapoint
$i$ of the mini-batch, respectively. The advantage of such a loss function is per-event
equivalence of the original and drone model, in addition to equivalence
of performance. For the drone training detailed in this article, standard
mini-batch stochastic gradient descent is used. A feature of this method
is that the drone classifier does not see the labels of the training data,
but rather learns the same properties from the original classifier.
This is therefore a neural network that learns from another neural network in an
empirical manner.

\subsection{Model morphing during the learning phase}

In order to keep the hyperparameter space to the minimum required level,
additional degrees of freedom are added only when required.
This removes the possibility of choosing an incorrect size of the
drone network. During the learning phase, the following conditions are required
to trigger the extension of the hidden layer in the $j^{\rm th}$ epoch:
\begin{align}
\delta_{j} &\equiv |\mathcal{L}_j-\mathcal{L}_{j-1}|/\mathcal{L}_j < \kappa,\label{eq:cond1}\\
\sigma_{j} &\equiv m (1 - e^{-b(\hat{t} + n)})\delta_{j}\mathcal{L}_j \nonumber\\
\mathcal{L}_j &< \hat{\mathcal{L}} - \sigma_{j} \label{eq:cond2},
\end{align}
where $\kappa$ is the required threshold, $\sigma$ is the required minimum improvement
of the loss function and $\hat{\mathcal{L}}$ is the value of the loss function when
the hidden layer was last extended. The required improvement starts from a minimum at $n$,
increases with epoch number after previous extension $\hat{t}$ and steepness $b$
until a maximum at $m$. The precise values of the parameters
$\kappa$, $n$, $m$, $b$ are not of particular importance. Rather, the topology described by
eqs.~\ref{eq:cond1} and \ref{eq:cond2} is crucial. The relative loss function improvement,
$\delta_{j}$, can never realistically be larger than $1$ and the limit, $\kappa$, at which
no significant improvement occurs is acceptably set at 0.02 (smaller than $2\sigma$
standard deviations). The descent in loss space, $\hat{\mathcal{L}} - \mathcal{L}_j$,
is further required to be significantly large, minimizing the chance of getting stuck in
isolated local minima. The function, $\sigma_{j}$, is chosen to increase this requirement
with each epoch for two reasons - it is bounded and can approach its asymptote arbitrarily fast.
It scales $\delta_{j}$ such that the loss descent must be significant
before an update is triggered. Since $\delta_{j}$ is expected to decrease with epoch number,
the minumum and maximum values of $\sigma_{j}$ are chosen as such:
\begin{align}
\sigma_{j}(\hat{t} = 0) &\equiv min(\sigma_{j}) \equiv 2.5\delta_{j}\mathcal{L}_j \implies 5\sigma ~\text{std.dev.} \\
\sigma_{j}(\hat{t} = \infty) &\equiv min(\sigma_{j}) \equiv 25\delta_{j}\mathcal{L}_j \implies 50\sigma ~\text{std.dev.}
\end{align}
The steepness, $b$, is chosen such that the transition from the minimum to maximum takes
on average 50 epochs. This ensures a change cannot be triggered immediately
after a previous one and the learning can still proceed if more freedom is indeed required.
Also, it allows the network to stabilize after a big change.

When the conditions in eqs.~\ref{eq:cond1} and \ref{eq:cond2} are met, the linear model
is updated to extend the weights matrices and bias vectors
to accommodate the layer addition.
The associated neurons are initialised with a zero weight
to ensure continuity of the loss function value.

\section{High energy physics applications}
\label{sec:hep}

\subsection{B physics}

\subsubsection{Data sample}

In order to demonstrate the functionality of the toolkit, data samples generated
from the RapidSim package~\cite{rapid} are used. The interesting signal is chosen
to be the $\Bs\to\jpsi(\to\mu\mu )\phi (\to\PK\PK )$ decay, and the background is
the $\Dz\to\pi\pi\pi\pi$ decay. A total of 10000 candidates is generated for each decay.

\subsubsection{Training of the original classifier}
\label{sec:orig_training}

The machine learning classifier, using the Keras framework~\cite{keras,adam},
is constructed as a locally connected first layer (in which filters are applied
to different regions in contrast to a full convolution layer), followed by a pooling layer,
and a standard dense layer. The exact definition can be found below.
\begin{lstlisting}
classifier = Sequential()
classifier.add(LocallyConnected1D(
filters = 90, kernel_size = 2,
activation = 'sigmoid',
input_shape = (len(setTrain[0]), 1)))
classifier.add(GlobalMaxPooling1D())
classifier.add(Dense(30, activation = 'sigmoid'))
classifier.add(Dense(1, activation = 'sigmoid'))
classifier.compile(optimizer = 'adam',
loss = 'binary_crossentropy'
, metrics = ['accuracy'])
\end{lstlisting}

\begin{figure*}[t]
\centering
\includegraphics[width=0.33\textwidth]{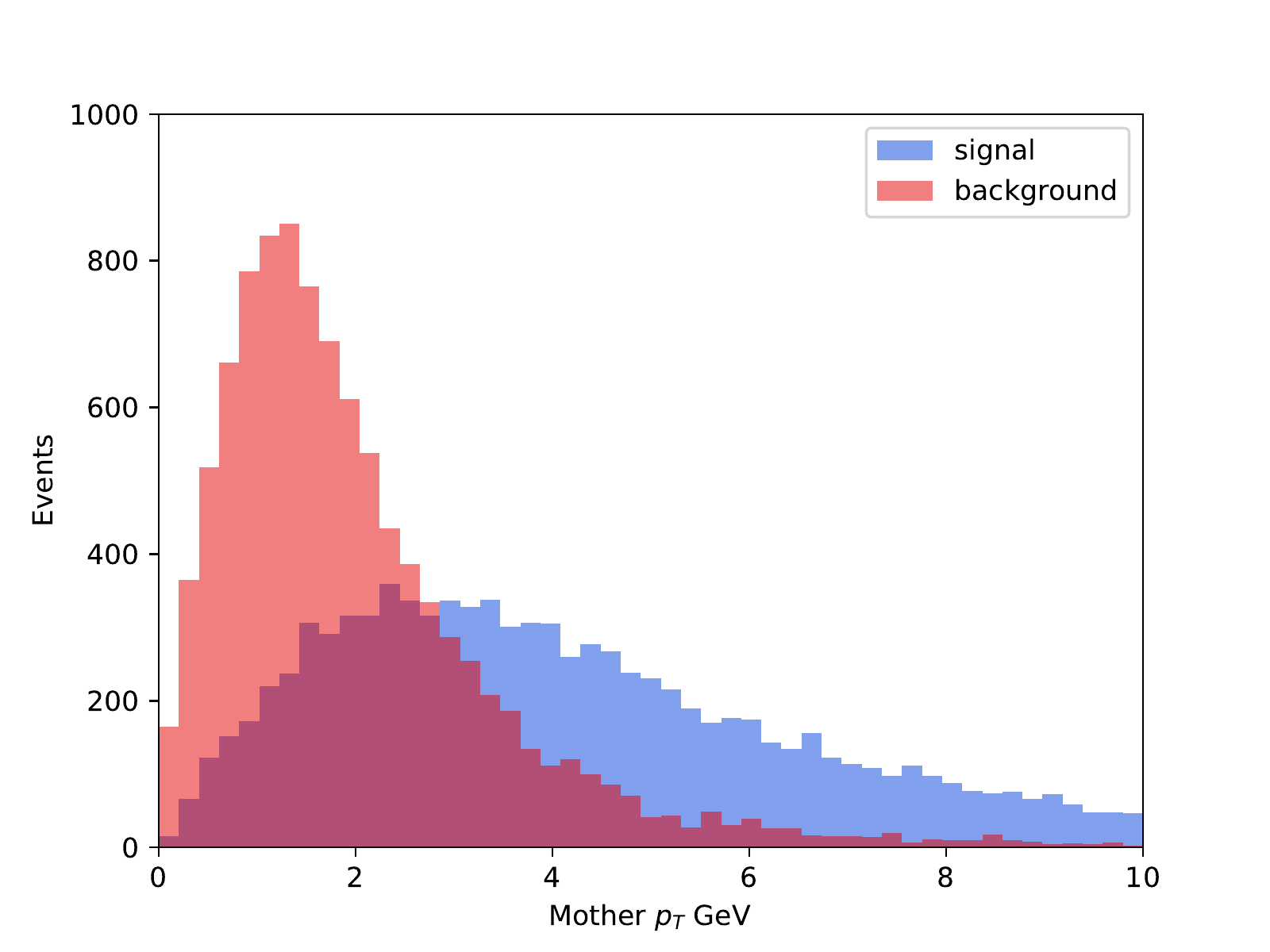}
\includegraphics[width=0.33\textwidth]{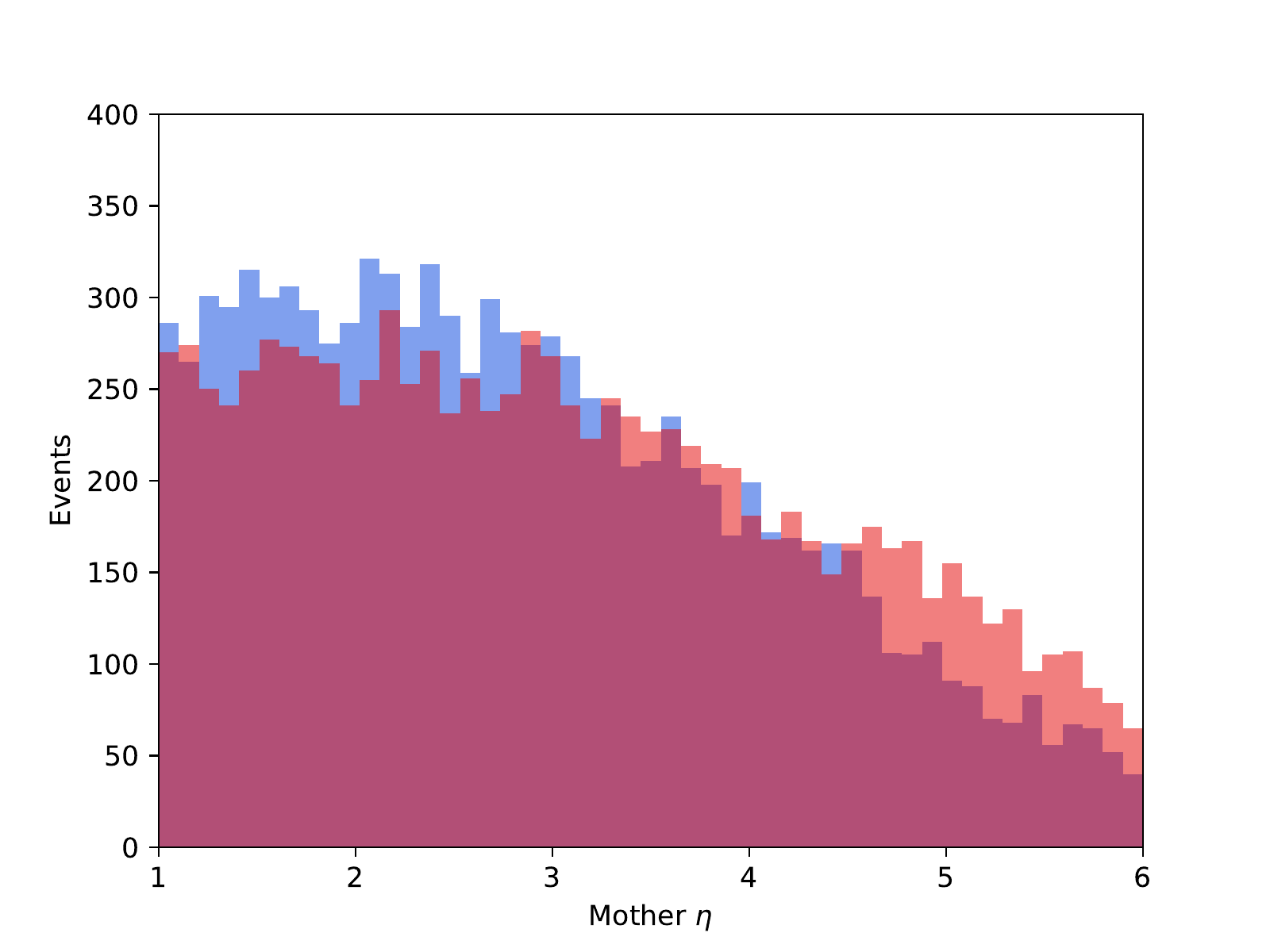}
\includegraphics[width=0.33\textwidth]{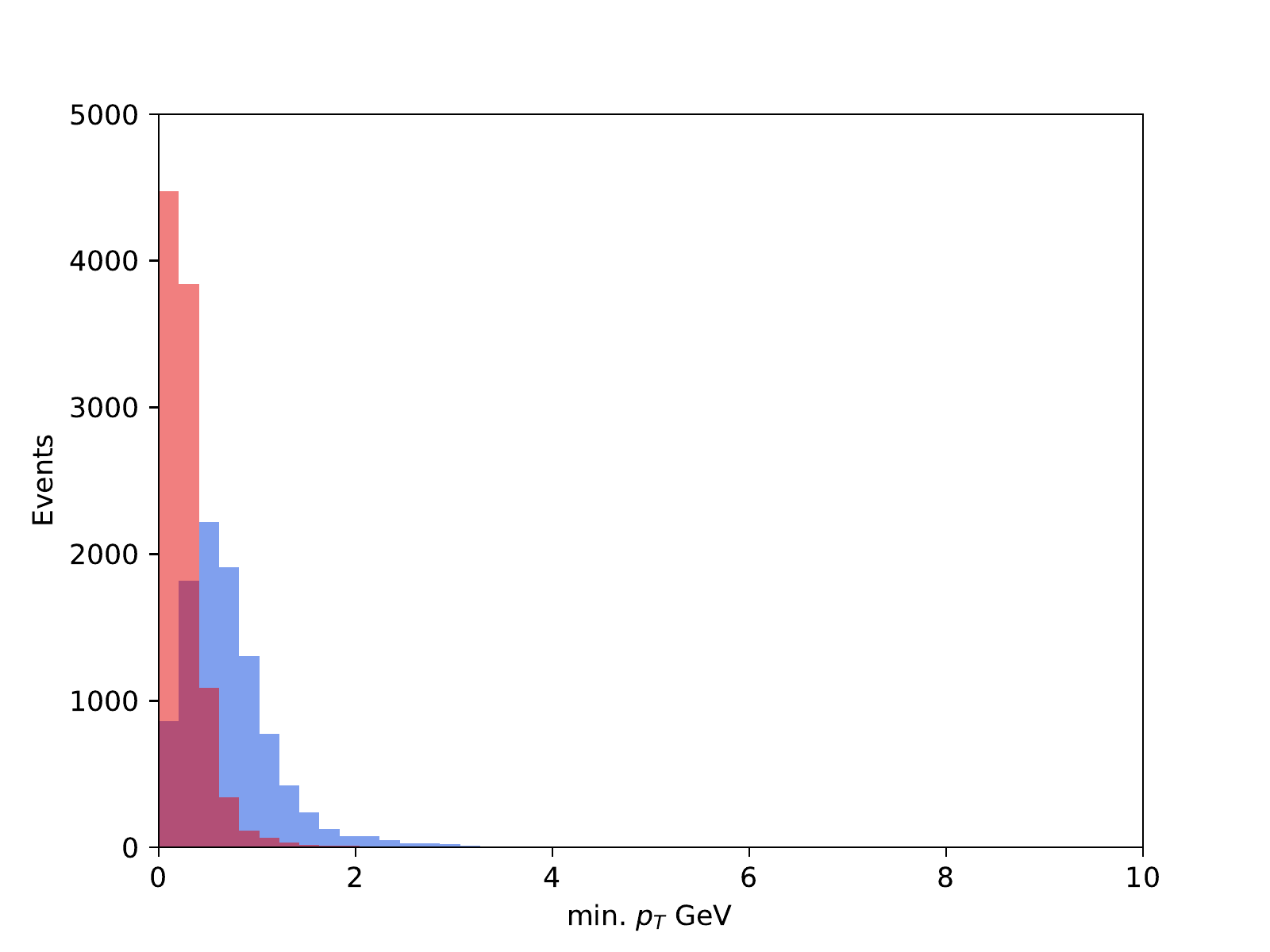}
\includegraphics[width=0.33\textwidth]{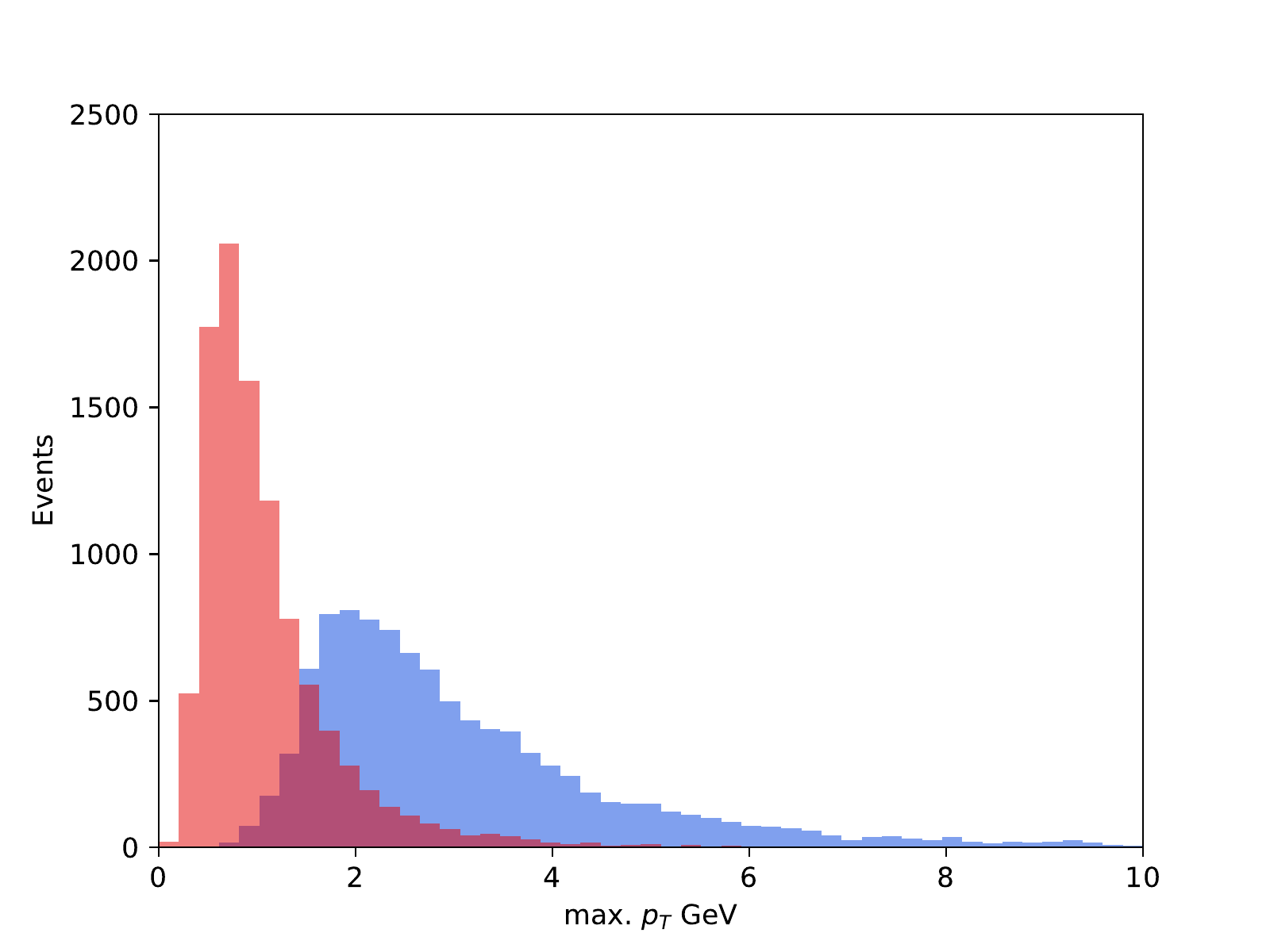}
\includegraphics[width=0.33\textwidth]{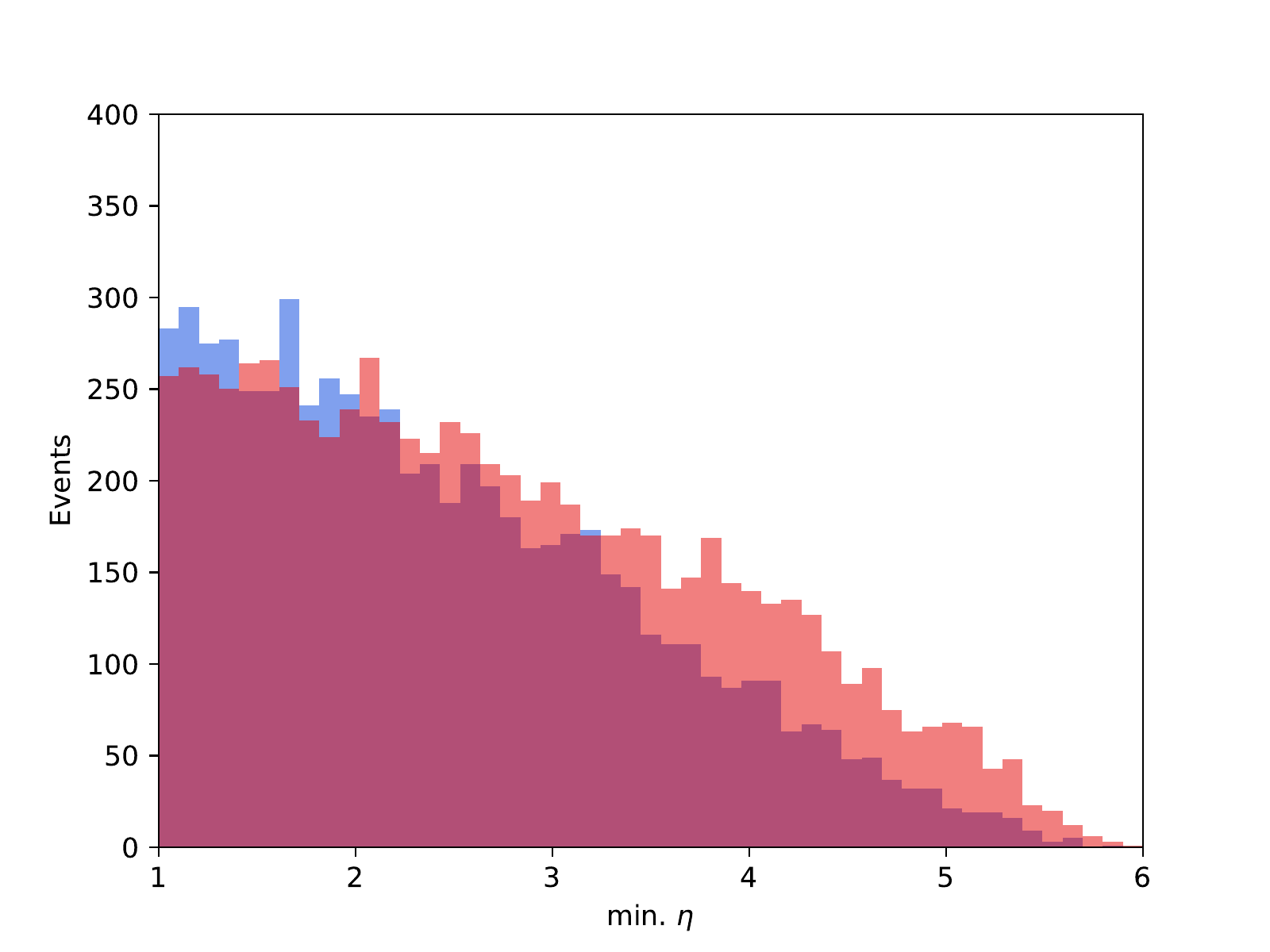}
\includegraphics[width=0.33\textwidth]{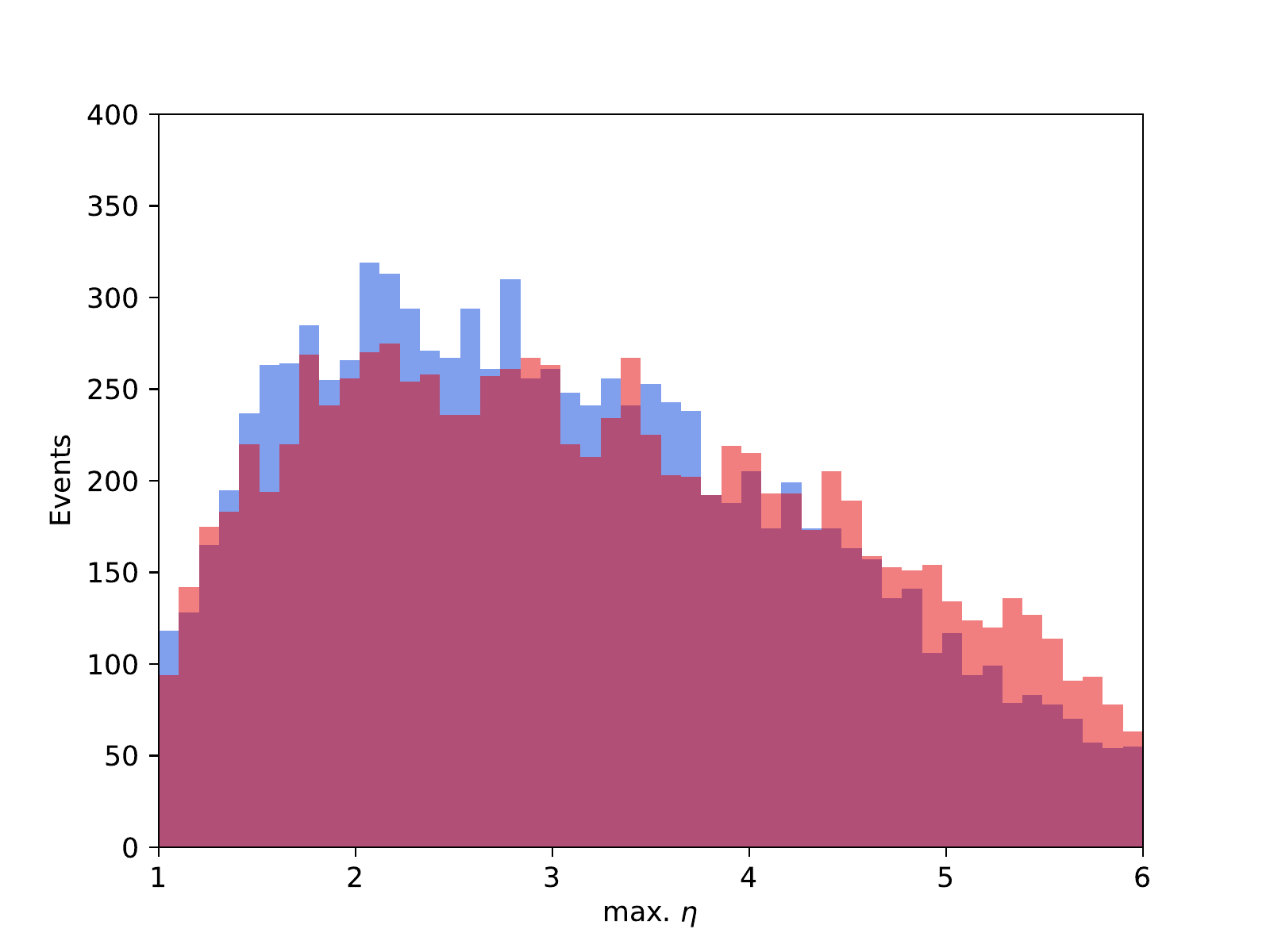}
\caption{\small Comparison of the signal and background distributions
used to train the Keras B decay classifier.}
\label{fig:inputs}
\end{figure*}

The neural network is trained using kinematic properties of the respective decays.
These include the pseudorapidity, $\eta$, and momentum transverse to the direction of the
input proton beams, \pt, of the decaying particle. In addition, the minimum and maximum \pt and $\eta$
of the final state particles is used. The signal and background distributions of the input variables
are shown in Fig.~\ref{fig:inputs}.

In the training of the original classifier, half of the data is
reserved in order to test for overtraining.

\subsection{Jet separation}
\label{sec:hepGPD}

\subsubsection{Data sample}

A further demonstration is provided demonstrating a classifiers ability to separate different
kinds of jets. The data sample to show this has been generated from Pythia~\cite{Sjostrand:2007gs}
simulating pp collisions at 14\tev.
The jets themselves are reconstructed in the Rivet analysis framework~\cite{Buckley:2010ar}
and are created using the FastJet~\cite{Cacciari:2011ma} package using the $K_t$ algorithm~\cite{Salam:2007xv}
(the definition
of the $K_t$ variable and a review of jet reconstruction algorithms
may be found in Refs~\cite{kt} and \cite{Atkin:2015msa} respectively).
A jet \pt requirement of 20\gev is imposed on all jets.
All other parameters remain at the default values for Rivet version 2.5.4.
The signal sample is chosen to correspond to a $qg\to Wq$ type of interaction,
whereas the background is chosen to correspond to a $gg \to gg$ type. These correspond
to the Rivet analyses named {\tt MC\_WJETS} and {\tt MC\_QCD}, respectively.
Jets that originate from gluons in the final state form a background to many
analyses, therefore efficient rejection of such processes is important in making
measurements~\cite{Komiske:2016rsd}.

\subsubsection{Training of the original classifier}

The machine learning classifier chosen is also a Keras-based convolutional neural net,
constructed in an similar way as described in Sec~\ref{sec:orig_training}
\begin{lstlisting}
classifier = Sequential()
classifier.add(LocallyConnected1D(
filters = 90, kernel_size = 2,
activation = 'sigmoid',
input_shape = (len(setTrain[0]), 1)))
classifier.add(GlobalMaxPooling1D())
classifier.add(Dense(30, activation = 'sigmoid'))
classifier.add(Dense(1, activation = 'sigmoid'))
classifier.compile(optimizer = 'adam',
loss = 'binary_crossentropy'
, metrics = ['accuracy'])
\end{lstlisting}

\begin{figure*}[t]
\centering
\includegraphics[width=0.33\textwidth]{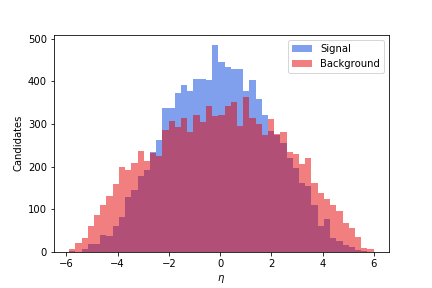}
\includegraphics[width=0.33\textwidth]{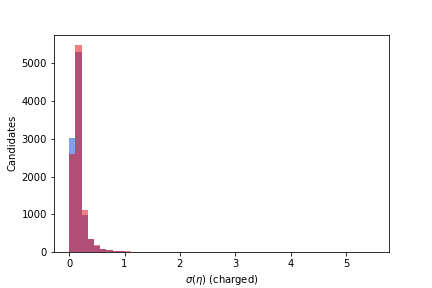}
\includegraphics[width=0.33\textwidth]{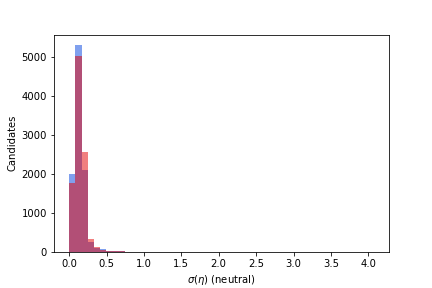}
\includegraphics[width=0.33\textwidth]{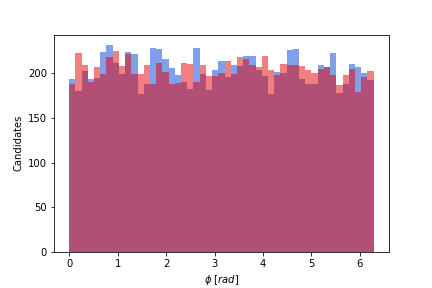}
\includegraphics[width=0.33\textwidth]{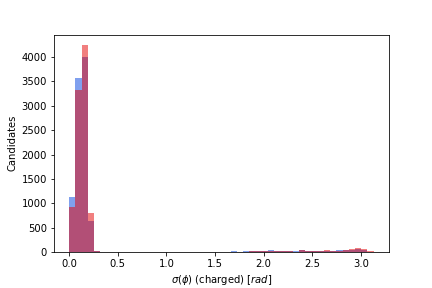}
\includegraphics[width=0.33\textwidth]{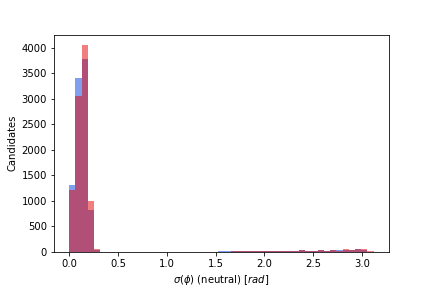}
\includegraphics[width=0.33\textwidth]{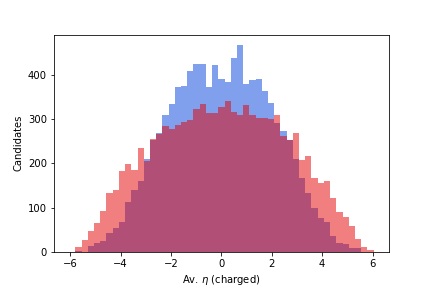}
\includegraphics[width=0.33\textwidth]{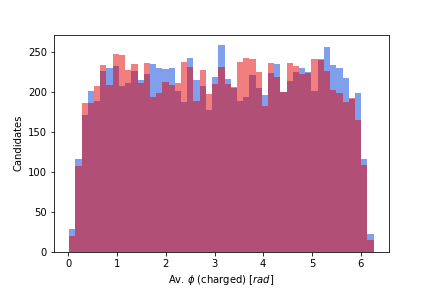}
\includegraphics[width=0.33\textwidth]{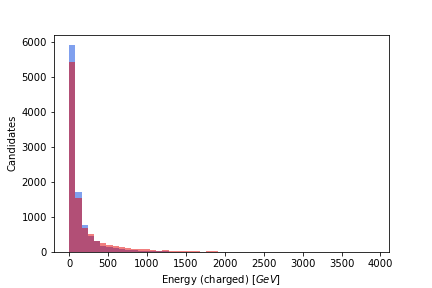}
\includegraphics[width=0.33\textwidth]{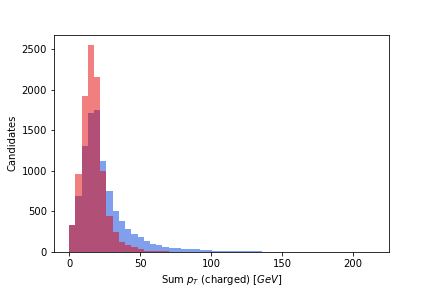}
\includegraphics[width=0.33\textwidth]{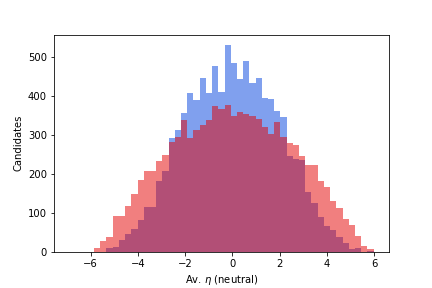}
\includegraphics[width=0.33\textwidth]{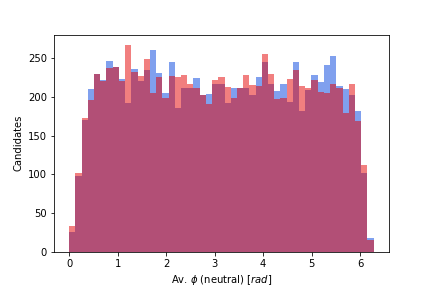}
\includegraphics[width=0.33\textwidth]{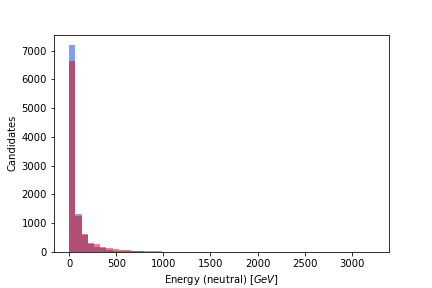}
\includegraphics[width=0.33\textwidth]{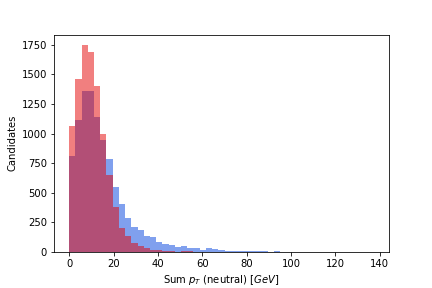}
\includegraphics[width=0.33\textwidth]{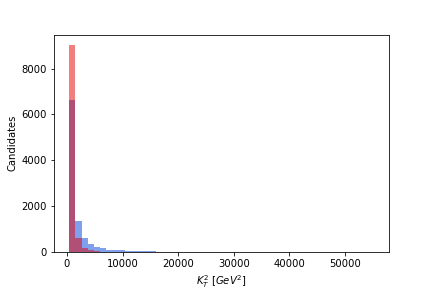}
\includegraphics[width=0.33\textwidth]{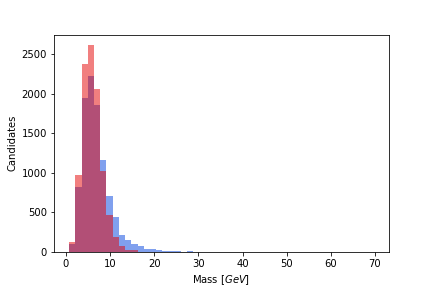}
\includegraphics[width=0.33\textwidth]{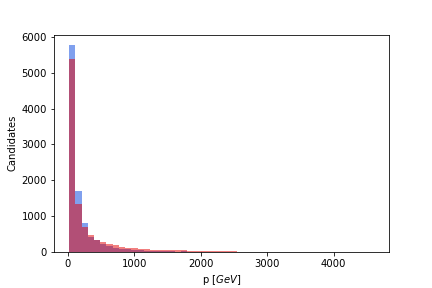}
\caption{\small Comparison of the signal and background distributions
used to train the Keras jet separation classifier.}
\label{fig:inputsGPD}
\end{figure*}

The training data is based around the properties of the measured jets. The list of features
taken consists of the azimuthal angle, $\phi$, $\eta$ of the jet; the spread of neutral and 
hadronic contributions to the jet in the $\phi$, $\eta$ variables, along with average and energy weighted
kinematic variables. In total 17 different features are used.
The signal and background distributions of the input variables
are shown in Fig.~\ref{fig:inputsGPD}.

\subsection{Drone conversions}

The drone neural networks are trained following the procedure outlined in Sec.~\ref{sec:dlearn},
In total, 300 epochs are used with
the learning rate of the stochastic gradient descent set to 0.05.
The value of $\kappa$ is chosen to be 0.02, the value of $b$ is chosen to
be 0.04 and the value of $m$ is chosen to be 50.

\begin{figure*}[t]
\centering
\includegraphics[width=0.45\textwidth]{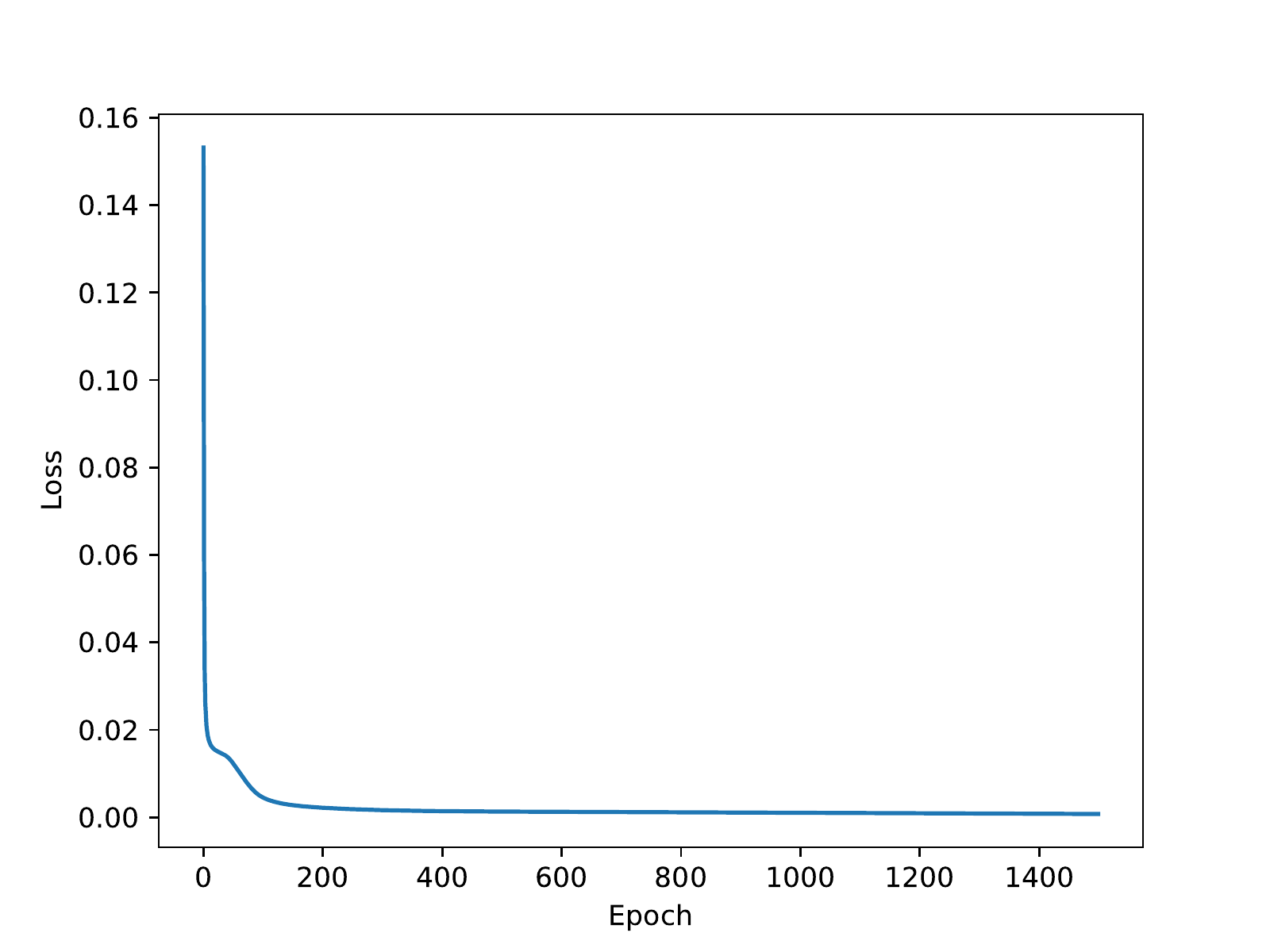}
\includegraphics[width=0.45\textwidth]{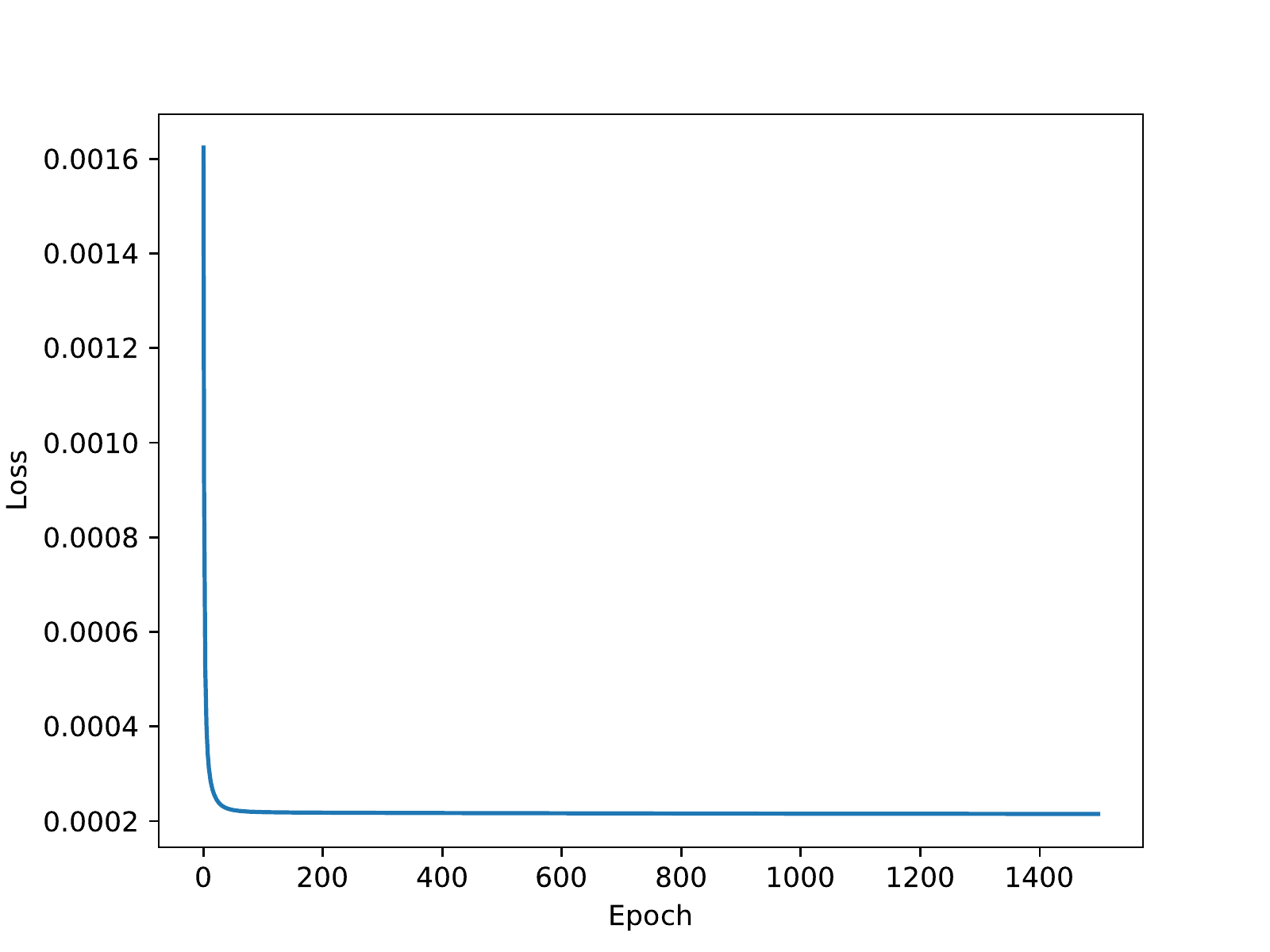}
\caption{\small
Convergence of the loss function during the drone training
  for the case of the B
  decay (left) and jet separation (right) examples.
}
\label{fig:loss}
\end{figure*}

\begin{figure*}[t]
\centering
\includegraphics[width=0.45\textwidth]{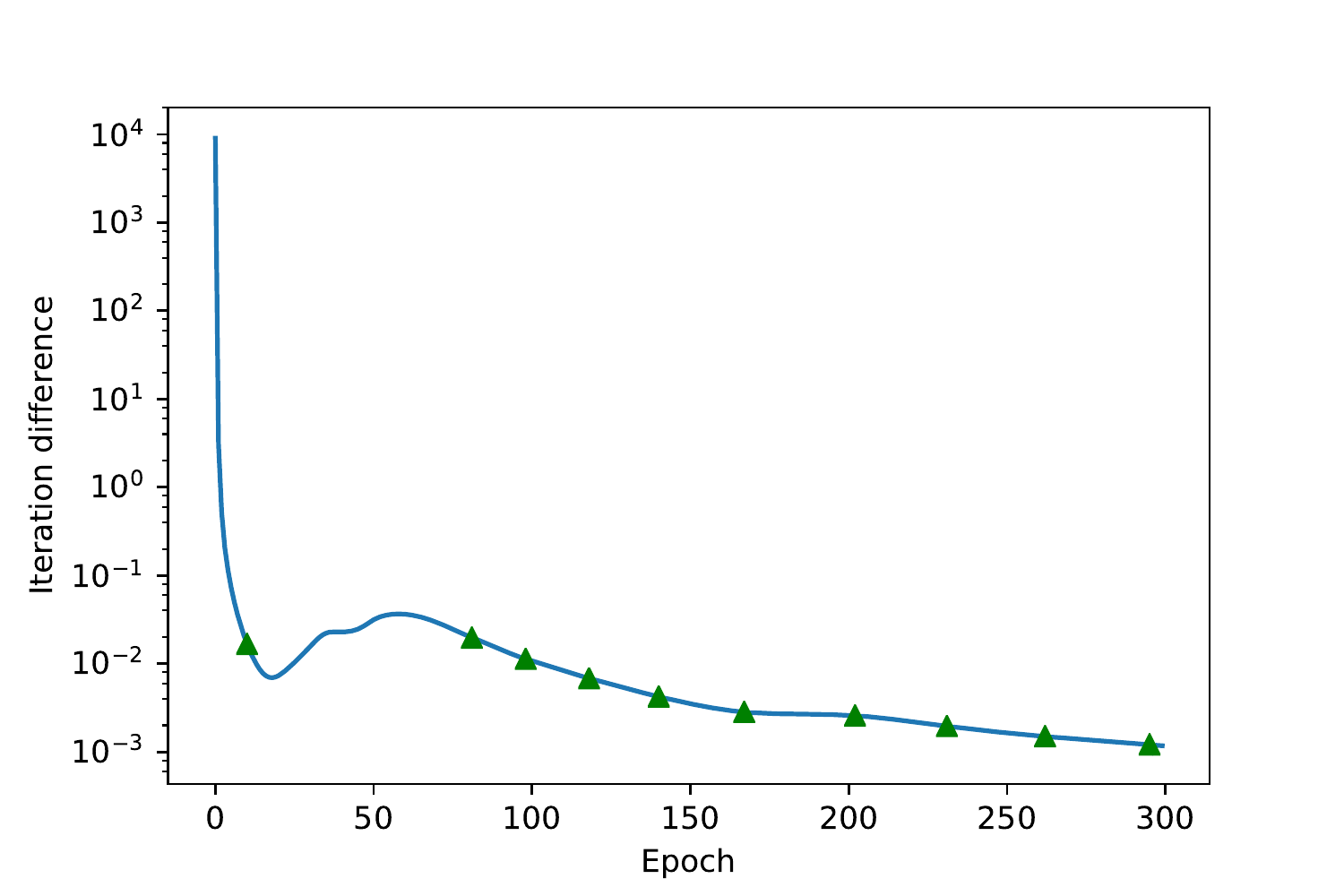}
\includegraphics[width=0.45\textwidth]{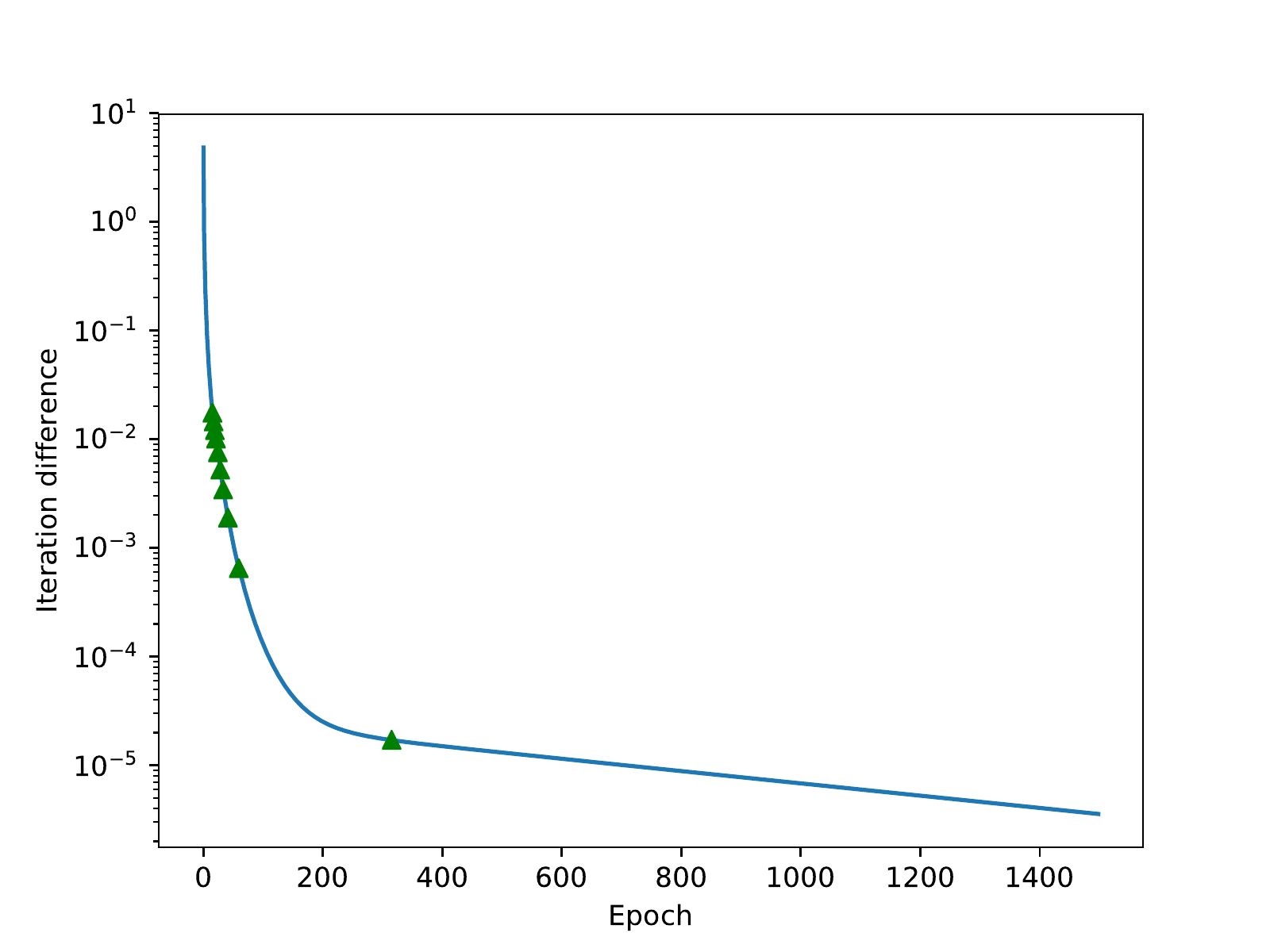}
\caption{\small
Difference in the loss function with respect to the previous iteration
  for the case of the B
  decay (left) and jet separation (right) examples.
  The green triangles
depict the epoch number in which the number of hyperperameters was increased.
}
\label{fig:iterdiff}
\end{figure*}

\begin{figure*}[t]
\centering
\includegraphics[width=0.45\textwidth]{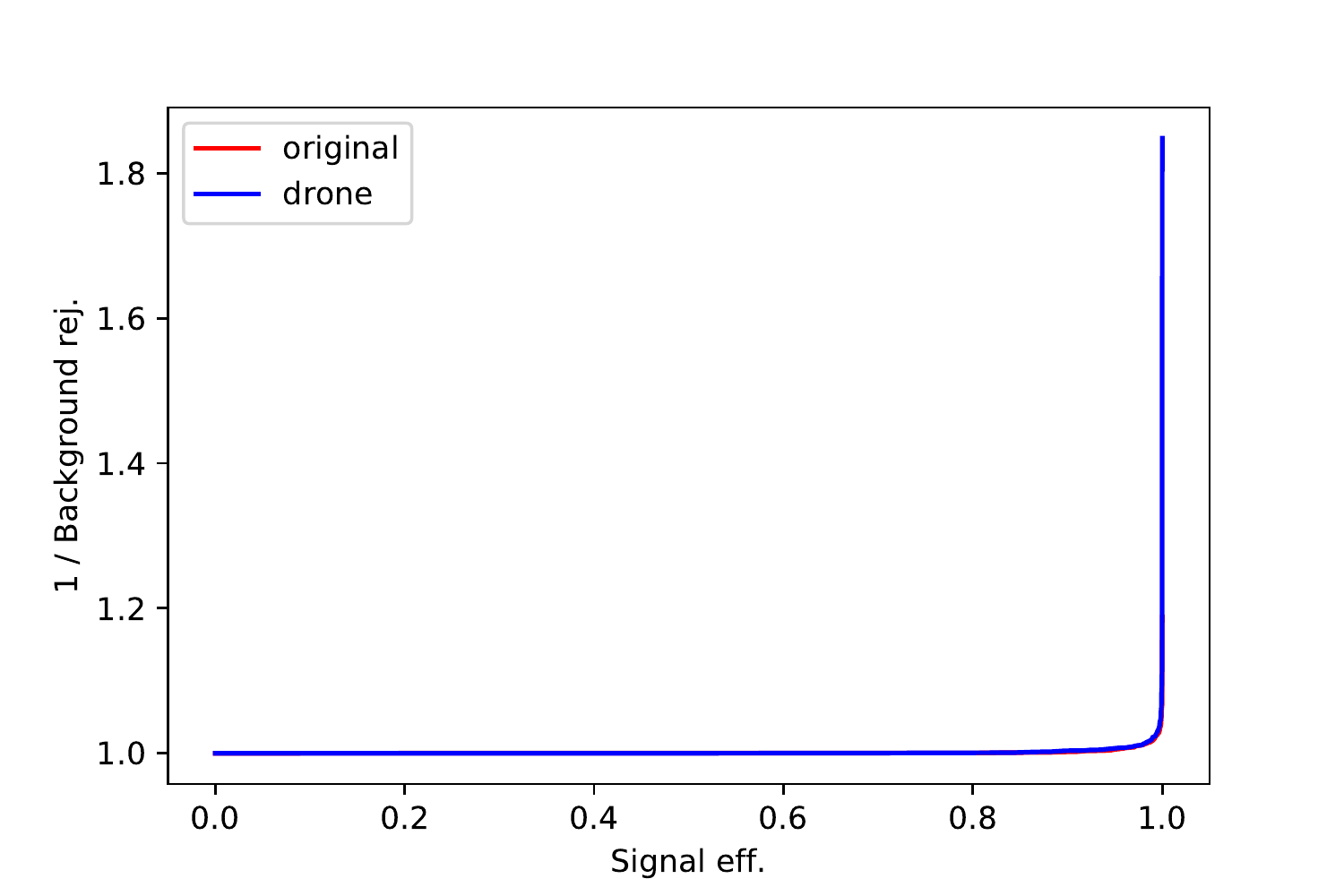}
\includegraphics[width=0.45\textwidth]{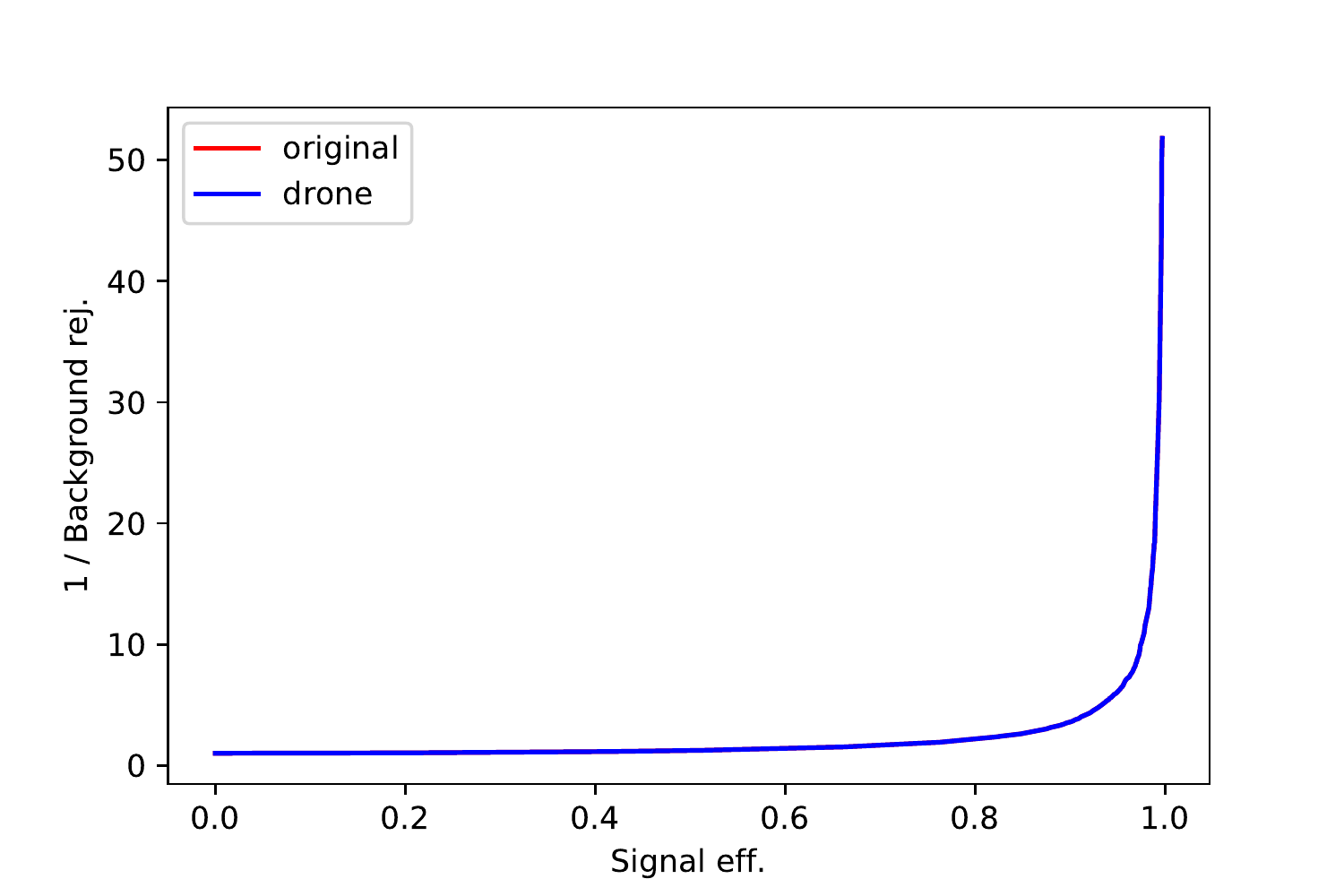}
\caption{\small
  Signal efficiency versus background rejection of the original classifier (red) and drone
  approximation (blue)
  for the case of the B
  decay (left) and jet separation (right) examples.
}
\label{fig:roc}
\end{figure*}

The loss history of the drone approximations are shown in Fig.~\ref{fig:loss}
as a function of epoch number.
The convergence is also shown in Fig.~\ref{fig:iterdiff}, which shows
the difference in the value of the loss function with respect to the previous
epoch. The epochs that trigger an increase in the number of hyperparameters
are also overlaid.
In total for the case of B decays and for the case of the jet separation classifier, 
an increase was triggered 10 times.
The total number
of parameters in the final drone neural networks are therefore 121 and 286 for the B decay drone
and the jet separation drone, respectively. It is interesting
to note that with the algorithm design of Sec.~\ref{sec:dlearn}, the introduction
of the new parameter space causes the drone networks to learn faster, as evidenced by
increases in Fig.~\ref{fig:iterdiff} with continuing descent of the loss functions.
The performance of the original classifiers compared to the drone classifiers are shown in Figure~\ref{fig:roc}.

\section{Drone storage and transferability and suitability
for low-latency environments}
\label{sec:storage}

The hyperparameters and structure of the drone are required to be
portable and easily stored for later usage. For this the {\tt JSON} format was chosen as
mediator. It is human-readable and easily accessible in the {\tt Python} and {\tt C++}
environments commonly used in HEP. Thus, it is readily deployable in both personal and production environments.

Provided is a tool to export and save a drone neural network to a {\tt JSON}
formatted file which preserves the input \& output structure,
the layers and nodes, all hyperparameters and activation functions.
The drone configuration is later read in by an equivalent tool into the production software framework,
which then constructs a class object based on the Keras model. The {\tt C++} class implements
a flexible member structure that is capable of completely reproducing the original drone. The production
implementation may be used for all data reduction levels, be it in the form of a low-latency trigger
for example up to the latest stages of data handling and output.

A major advantage of this method is that analysts and users have the full freedom of latest developments
of industry standards, but need only to support a more manageable implementation in the low-latency
software. This is further aided by projects such as ONNX~\cite{ONNX}, which enable classifiers from a wider
range of software packages to be converted to a framework in which an approximation converter
is available.

The identical performance show in Fig.~\ref{fig:roc} is clearly the ideal scenario, even though
such good agreement is not always required to give better results than other low-latency methods.
However it is worth noting that the drones created in the examples of Sec.~\ref{sec:hep} are faster to
evaluate. The comparison of the time taken for each model evaluation, determined from a desktop
using a Intel Core i5-7267U processor is shown in Table~\ref{tab:comp}.
\begin{table}[t]
  \centering
  \caption{Hyperparameter number comparisons of the original models and drone
  approximations for the HEP examples. \label{tab:comp_param}}
  \begin{tabular}{l|rr}
                   & original model                  & drone \\
    \hline
    B decay        & 4,111 & 121 \\
    jet separation & 7081 & 286 \\
  \end{tabular}
\end{table}

\begin{table}[t]
  \centering
  \caption{Processing time comparisons of the original models and drone
  approximations for the HEP examples. \label{tab:comp}}
  \begin{tabular}{l|rr}
                   & original model                  & drone \\
    \hline
    B decay        & $3.87 \times 10^{-4}$ s & $4.8 \times 10^{-5}$ s \\
    jet separation & $4.79 \times 10^{-4}$ s & $6.2 \times 10^{-5}$ s \\
  \end{tabular}
\end{table}

\section{Summary}
\label{sec:summary}

It has been demonstrated that for the case of a high energy physics
event selection application, a drone neural network is able to accurately
approximate and learn the features of a neural network with a different
structure. The proposed algorithm design allows the drone to learn the
aforementioned features without ever having access to the training data,
or indeed any data, but only with appropriate questioning of the original model.

The equivalency of the outputs of the drone and original model enables an
analyst to treat both the original and the drone in the same way. The creation
of a drone in a standardised form permits an analyst to use any desired machine-learning
package to isolate a decay signature, and from this create a classifier
guaranteed to be suitable for execution in the {\tt C++} real-time data selection frameworks.

\section*{Acknowledgements}

\noindent
We acknowledge support from
the NWO (The Netherlands) and STFC (United Kingdom).
We are indebted to the communities behind the multiple open
source software packages on which we depend.
This project has received funding from the European Union’s Horizon
2020 research and innovation programme under the Marie Skłodowska-Curie
grant agreement No 676108.


\begin{thebibliography}{10}
\expandafter\ifx\csname url\endcsname\relax
  \def\url#1{\texttt{#1}}\fi
\expandafter\ifx\csname urlprefix\endcsname\relax\def\urlprefix{URL }\fi
\expandafter\ifx\csname href\endcsname\relax
  \def\href#1#2{#2} \def\path#1{#1}\fi

\bibitem{Alves:2008zz}
A.~A. Alves~Jr., et~al., {The \lhcb detector at the LHC}, JINST 3 (2008)
  S08005.
\newblock \href {http://dx.doi.org/10.1088/1748-0221/3/08/S08005}
  {\path{doi:10.1088/1748-0221/3/08/S08005}}.

\bibitem{LHCb-DP-2014-002}
R.~Aaij, et~al., {LHCb detector performance}, Int. J. Mod. Phys. A30 (2015)
  1530022.
\newblock \href {http://arxiv.org/abs/1412.6352} {\path{arXiv:1412.6352}},
  \href {http://dx.doi.org/10.1142/S0217751X15300227}
  {\path{doi:10.1142/S0217751X15300227}}.

\bibitem{Xu:2016mik}
Z.~Xu, M.~Tobin, {Novel real-time alignment and calibration of the LHCb
  detector in Run II}, Nucl. Instrum. Meth. A824 (2016) 70--71.
\newblock \href {http://dx.doi.org/10.1016/j.nima.2015.11.040}
  {\path{doi:10.1016/j.nima.2015.11.040}}.

\bibitem{Aaij:2016rxn}
R.~Aaij, et~al., {Tesla : an application for real-time data analysis in High
  Energy Physics}, Comput. Phys. Commun. 208 (2016) 35--42.
\newblock \href {http://arxiv.org/abs/1604.05596} {\path{arXiv:1604.05596}},
  \href {http://dx.doi.org/10.1016/j.cpc.2016.07.022}
  {\path{doi:10.1016/j.cpc.2016.07.022}}.

\bibitem{Hocker:2007ht}
A.~Hocker, et~al., {TMVA - Toolkit for Multivariate Data Analysis}, PoS ACAT
  (2007) 040.
\newblock \href {http://arxiv.org/abs/physics/0703039}
  {\path{arXiv:physics/0703039}}.

\bibitem{Feindt:2006pm}
M.~Feindt, U.~Kerzel, {The NeuroBayes neural network package}, Nucl. Instrum.
  Meth. A559 (2006) 190--194.
\newblock \href {http://dx.doi.org/10.1016/j.nima.2005.11.166}
  {\path{doi:10.1016/j.nima.2005.11.166}}.

\bibitem{Pedregosa:2012toh}
F.~Pedregosa, et~al., {Scikit-learn: Machine Learning in Python}, J. Machine
  Learning Res. 12 (2011) 2825--2830.
\newblock \href {http://arxiv.org/abs/1201.0490} {\path{arXiv:1201.0490}}.

\bibitem{keras}
F.~Chollet, et~al., Keras, \url{https://github.com/fchollet/keras} (2015).

\bibitem{Barrand:2001ny}
G.~Barrand, et~al., {GAUDI - A software architecture and framework for building
  HEP data processing applications}, Comput. Phys. Commun. 140 (2001) 45--55.
\newblock \href {http://dx.doi.org/10.1016/S0010-4655(01)00254-5}
  {\path{doi:10.1016/S0010-4655(01)00254-5}}.

\bibitem{Gligorov:2012qt}
V.~V. Gligorov, M.~Williams, {Efficient, reliable and fast high-level
  triggering using a bonsai boosted decision tree}, JINST 8 (2013) P02013.
\newblock \href {http://arxiv.org/abs/1210.6861} {\path{arXiv:1210.6861}},
  \href {http://dx.doi.org/10.1088/1748-0221/8/02/P02013}
  {\path{doi:10.1088/1748-0221/8/02/P02013}}.

\bibitem{losssurfaces}
A.~Choromanska, M.~Henaff, M.~Mathieu, G.~B. Arous, Y.~LeCun,
  \href{http://arxiv.org/abs/1412.0233}{The loss surface of multilayer
  networks}, CoRR abs/1412.0233.
\newblock \href {http://arxiv.org/abs/1412.0233} {\path{arXiv:1412.0233}}.
\newline\urlprefix\url{http://arxiv.org/abs/1412.0233}

\bibitem{HORNIK1991251}
K.~Hornik,
  \href{http://www.sciencedirect.com/science/article/pii/089360809190009T}{Approximation
  capabilities of multilayer feedforward networks}, Neural Networks 4~(2)
  (1991) 251 -- 257.
\newblock \href {http://dx.doi.org/10.1016/0893-6080(91)90009-T}
  {\path{doi:10.1016/0893-6080(91)90009-T}}.
\newline\urlprefix\url{http://www.sciencedirect.com/science/article/pii/089360809190009T}

\bibitem{rapid}
G.~A. Cowan, D.~C. Craik, M.~D. Needham, {RapidSim: an application for the fast
  simulation of heavy-quark hadron decays}, Comput. Phys. Commun. 214 (2017)
  239--246.
\newblock \href {http://arxiv.org/abs/1612.07489} {\path{arXiv:1612.07489}},
  \href {http://dx.doi.org/10.1016/j.cpc.2017.01.029}
  {\path{doi:10.1016/j.cpc.2017.01.029}}.

\bibitem{adam}
D.~P. Kingma, J.~Ba, \href{http://arxiv.org/abs/1412.6980}{Adam: {A} method for
  stochastic optimization}, CoRR abs/1412.6980.
\newblock \href {http://arxiv.org/abs/1412.6980} {\path{arXiv:1412.6980}}.
\newline\urlprefix\url{http://arxiv.org/abs/1412.6980}

\bibitem{Sjostrand:2007gs}
T.~Sj\"{o}strand, S.~Mrenna, P.~Skands, {A brief introduction to PYTHIA 8.1},
  Comput. Phys. Commun. 178 (2008) 852--867.
\newblock \href {http://arxiv.org/abs/0710.3820} {\path{arXiv:0710.3820}},
  \href {http://dx.doi.org/10.1016/j.cpc.2008.01.036}
  {\path{doi:10.1016/j.cpc.2008.01.036}}.

\bibitem{Buckley:2010ar}
A.~Buckley, J.~Butterworth, L.~Lonnblad, D.~Grellscheid, H.~Hoeth, J.~Monk,
  H.~Schulz, F.~Siegert, {Rivet user manual}, Comput. Phys. Commun. 184 (2013)
  2803--2819.
\newblock \href {http://arxiv.org/abs/1003.0694} {\path{arXiv:1003.0694}},
  \href {http://dx.doi.org/10.1016/j.cpc.2013.05.021}
  {\path{doi:10.1016/j.cpc.2013.05.021}}.

\bibitem{Cacciari:2011ma}
M.~Cacciari, G.~P. Salam, G.~Soyez, {FastJet User Manual}, Eur. Phys. J. C72
  (2012) 1896.
\newblock \href {http://arxiv.org/abs/1111.6097} {\path{arXiv:1111.6097}},
  \href {http://dx.doi.org/10.1140/epjc/s10052-012-1896-2}
  {\path{doi:10.1140/epjc/s10052-012-1896-2}}.

\bibitem{Salam:2007xv}
G.~P. Salam, G.~Soyez, {A Practical Seedless Infrared-Safe Cone jet algorithm},
  JHEP 05 (2007) 086.
\newblock \href {http://arxiv.org/abs/0704.0292} {\path{arXiv:0704.0292}},
  \href {http://dx.doi.org/10.1088/1126-6708/2007/05/086}
  {\path{doi:10.1088/1126-6708/2007/05/086}}.

\bibitem{kt}
S.~Catani, Y.~L. Dokshitzer, M.~H. Seymour, B.~R. Webber, {Longitudinally
  invariant $K_t$ clustering algorithms for hadron hadron collisions}, Nucl.
  Phys. B406 (1993) 187--224.
\newblock \href {http://dx.doi.org/10.1016/0550-3213(93)90166-M}
  {\path{doi:10.1016/0550-3213(93)90166-M}}.

\bibitem{Atkin:2015msa}
R.~Atkin, {Review of jet reconstruction algorithms}, J. Phys. Conf. Ser.
  645~(1) (2015) 012008.
\newblock \href {http://dx.doi.org/10.1088/1742-6596/645/1/012008}
  {\path{doi:10.1088/1742-6596/645/1/012008}}.

\bibitem{Komiske:2016rsd}
P.~T. Komiske, E.~M. Metodiev, M.~D. Schwartz, {Deep learning in color: towards
  automated quark/gluon jet discrimination}, JHEP 01 (2017) 110.
\newblock \href {http://arxiv.org/abs/1612.01551} {\path{arXiv:1612.01551}},
  \href {http://dx.doi.org/10.1007/JHEP01(2017)110}
  {\path{doi:10.1007/JHEP01(2017)110}}.

\bibitem{ONNX}
\href{http://onnx.ai}{{ONNX: Open Neural Network Exchange Format}}.
\newline\urlprefix\url{http://onnx.ai}

\end{thebibliography}
\end{document}